# I AM 4 VHO: NEW APPROACH TO IMPROVE SEAMLESS VERTICAL HANDOVER IN HETEROGENEOUS WIRELESS NETWORKS

Omar Khattab[1] and Omar Alani[2]
School of Computing, Science & Engineering
University of Salford, UK

o.khattab@edu.salford.ac.uk
o.y.k.alani@salford.ac.uk

## ABSTRACT

*Two mechanisms have been proposed independently by IEEE and 3GPP; namely, Media Independent Handover (MIH) and Access Network Discovery and Selection Function (ANDSF), respectively. These mechanisms enable a seamless Vertical Handover (VHO) between the different types of technologies (3GPP and non-3GPP), such as GSM (Global System for Mobile Communication), Wireless Fidelity (Wi-Fi), Worldwide Interoperability for Microwave Access (WiMAX), Universal Mobile Telecommunications System (UMTS) and Long Term Evolution (LTE). In this paper, we overview these mechanisms and show their components, benefits and drawbacks. Then we present our Imperative Alternative MIH for Vertical Handover (I AM 4 VHO) approach based on the approaches that have been studied in the literature with better performance (packet loss and latency), less connection failure (probability of reject sessions), less complexity and more exhaustive for enhancing VHO heterogeneous wireless networks environment.*

## KEYWORDS

*Vertical Handover (VHO), Media Independent Handover (MIH), ANDSF (Access Network Discovery and Selection Function), Heterogeneous Wireless Networks.*

## 1. INTRODUCTION

With the advancement of wireless communication and computer technologies, mobile communication has been providing more versatile, portable and affordable networks services than ever. Therefore, the number of mobile users (MUs) communication networks has increased rapidly as an example; it has been reported that "today, there are billions of mobile phone subscribers, close to five billion people with access to television, and tens of millions of new internet users every year" [1] and there is a growing demand for services over broadband wireless networks due to diversity of services which can't be provided with a single wireless network anywhere anytime [2]. This fact means that heterogeneous environment of wireless networks, such as GSM (Global System for Mobile Communication), Wireless Fidelity (Wi-Fi), Worldwide Interoperability for Microwave Access (WiMAX), Universal Mobile Telecommunications System (UMTS) and Long Term Evolution (LTE) will coexist providing MU with roaming capability across different wireless networks. One of the challenging issues in Next Generation Wireless Systems (NGWS) is seamless Vertical Handover (VHO) during the mobility between these technologies; therefore, the telecommunication operators will be required to develop a strategy for interoperability of these different types of existing wireless networks to get the best connection anywhere anytime. To fulfill these requirements of seamless VHO two mechanisms have been proposed independently by IEEE and 3GPP [3]; namely, Media Independent Handover (MIH) and Access Network Discovery and Selection Function (ANDSF), respectively. Each of them enables a seamless VHO between the different types of technologies (3GPP and non-3GPP), such as UMTS, Wi-Fi, WiMAX and LTE, this is shown in Fig.1.

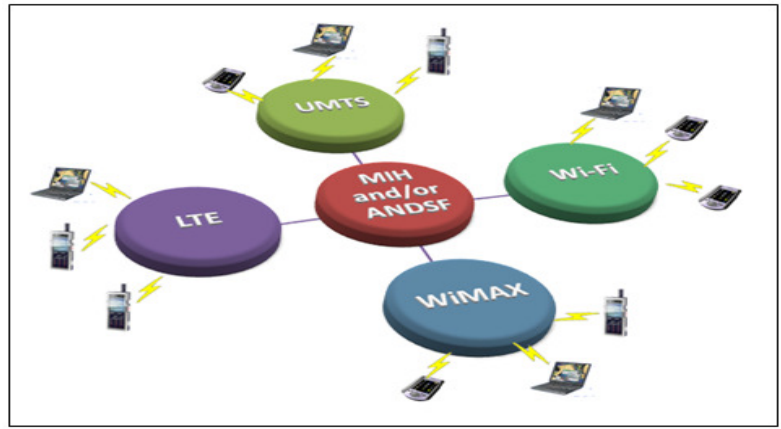

Figure 1: Various Radio Access Technologies (RATs) integration supported by MIH /ANDSF

## 1.1 Vertical Handover Procedure

The handover is a mechanism which allows the MUs to continue their ongoing sessions when moving within the same RAT (Radio Access Technology) coverage areas or traversing different RATs, is named Horizontal Handover (HHO) and VHO, respectively. In the literature most of the research papers divided VHO into three phases: Collecting Information, Decision and Execution, e.g. [4, 5] as described below.

**Handover Collecting Information**

In this phase, all the required information for VHO decision is gathered some related to the user preferences (such as cost, security), network (such as latency, coverage) and terminal (such as battery, velocity).

**Handover Decision**

In this phase, select the best RAT based on aforementioned information and informs the handover execution about that.

**Handover Execution**

In this phase, the active session for the MU will be maintained and continued on the new RAT; after that, the resources of old RAT is released eventually.

In this paper, we propose our Imperative Alternative MIH for Vertical Handover (I AM 4 VHO) approach based on the approaches that have been studied in the literature with better performance (packet loss and latency), less connection failure (probability of reject sessions), less complexity and more exhaustive for enhancing VHO heterogeneous wireless networks environment. The rest of the paper is organized as follows: section 2 presents previous work in heterogeneous integration approaches. In section 3, we present MIH and ANDSF mechanisms with their components and benefits as well as highlight drawbacks of combining between these mechanisms. In section 4, we present our Imperative Alternative MIH for Vertical Handover (I AM 4 VHO) approach and finally, we conclude the paper in section 5.

## 2. RELATED WORK

The key of VHO management procedure is the Access Network Selection (ANS) in the decision phase. There are many proposals proposed by researchers about ANS, e.g. [4, 5]; however, the proposed ANS schemes lack unity, while a number of issues still need to be resolved such as the goals discrepancy between user centric and network centric, where in user centric scheme the goal is how to get best connection anywhere anytime regardless of network operation complexities associated with this which matters from operator's perspective [6]. Although there are many approaches that have been presented to provide seamless VHO [2, 3, 11-16], we will consider three of them [3, 11 and 12] which have been evolved from each other such that the recent approach of them [3] combined between MIH and ANDSF mechanisms.

## 3. MIH AND ANDSF MECHANISMS

One challenge of wireless networks integration is the ubiquitous wireless access abilities which provide the seamless handover for any moving device in the heterogeneous networks. This challenge is important as MUs are becoming increasingly demanding for services regardless of the technological complexities associated with it. To fulfill these requirements of seamless handover two mechanisms have been proposed independently by IEEE and 3GPP; namely, MIH and ANDSF, respectively. Each of them enables seamless VHO between different types of technologies (3GPP and non-3GPP), such as UMTS, Wi-Fi and WiMAX. In this section, we overview MIH and ANDSF components and benefits as well as highlight drawbacks of combining between these mechanisms.

### 3.1 Media Independent Handover (MIH)

The IEEE group has proposed IEEE 802.21 standard Media Independent Handover (MIH) to provide a seamless VHO between different RATs. The MIH defines two entities: first, Point of Service (PoS) which is responsible for establishing communication between the network and MU under MIH and second, Point of Attachment (PoA) which is the RAT access point. Also MIH provides three main services: Media Independent Event Service (MIES), Media Independent Command Service (MICS) and Media Independent Information Service (MIIS).

**Media Independent Event Service (MIES)**

It is responsible for reporting the events after detecting, e.g. link up on the connection (established), link down (broken), link going down (breakdown imminent), link handover imminent, link handover complete, etc. [7].

**Media Independent Information Service (MIIS)**

It is responsible for collecting all information required to identify if a handover is needed or not and provide them to MU, e.g. available networks, locations, capabilities, cost, operator ID, etc. [7], this is shown in Fig.2.

**Media Independent Command Service (MICS)**

It is responsible for issuing the commands based on the information which is gathered by MIIS and MIES, e.g. MIH handover initiate, MIH handover prepare, etc. [7].

However, no handover decision is made within MIH [3], as the implementation of the decision of algorithm is out of the scope of MIH [8], the algorithms to be implemented are left to the designers [9], besides, there is inconsistency in MIH operation; hence, it needs some improvements [2].

### 3.2 Access Network Discovery and Selection Function (ANDSF)

The 3GPP group has proposed ANDSF to provide a seamless VHO between different RATs and to mitigate the impacts of radio signals impairment between 3GPP and non 3GPP. In this mechanism, there is no need to the measurements reports between the different RATs, and hence, no need to the modification on legacy radio systems (no additional cost).The ANDSF also works as store of RATs information that is queried by MU to make handover decision. This information about neighbor cells, operator's policies and preferences, QoS, capabilities, etc. [10], this is shown in Fig.3. In [11], new logical element proposed named Forward Authentication Function (FAF), it was collocated with the ANDSF and located in the target network. FAF plays the role of target RAT to perform its functionalities, e.g. if the MU moves toward 3GPP E-UTRAN, the FAF emulates Node-B, while if the MU moves toward WiMAX, the FAF emulates WiMAX Base Station (BS). The FAF has two main goals: first, to enable the transmission from WiMAX to 3GPP (Authentication). Second, to avoid direct link between 3GPP and WiMAX, i.e."avoid the WiMAX access scheduling measurement opportunities to the MU in order to measure neighbor 3GPP sites" [11]. Nevertheless, the authors in [11] lacked to tackle two vital aspects in the VHO procedure: first, the source network was not informed by MU about its movement to the target network which resulted in packet losses and second, it lacked releasing procedure for the resources of the network. In [12], the Data Forwarding Function (DFF) logical entity located in source network was proposed to solve the problems that were raised in [11].

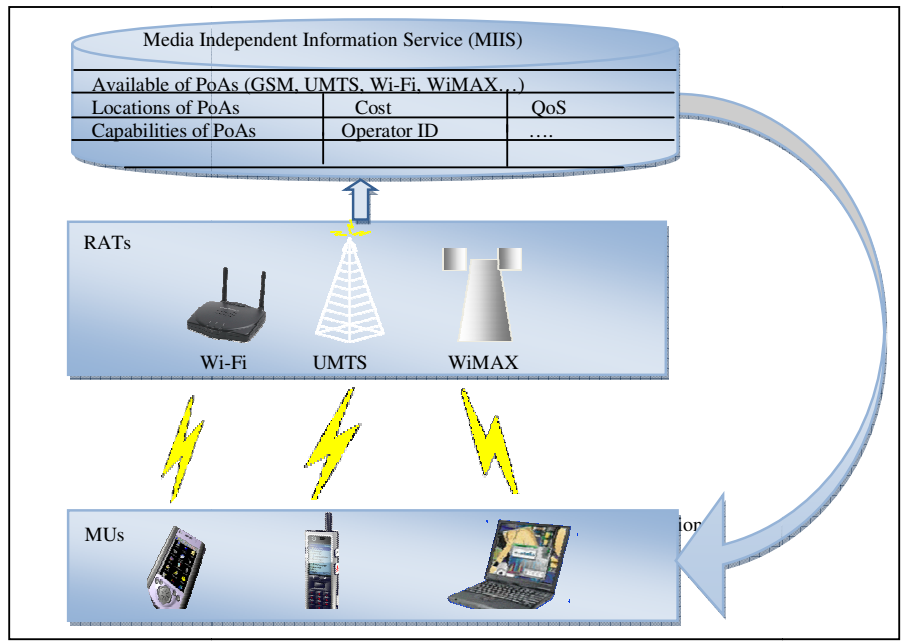

Figure 2: Media Independent Information Service (MIIS) passing information
about Radio Access Technologies (RATs) to Mobile Users (MUs)

## 3.3 Drawbacks with MIH and ANDSF Combination

In [3], combination between MIH and ANDSF was proposed; hence, there was no need for FAF and DFF to be existing. In addition for this combination, the MU obtained operator's policies from the ANDSF which has the role of selecting the target network. However, the authors in [3] have not provided evaluations or validations of their work which was not exhaustive and complex as a result of the combination between two mechanisms.
From the above discussion we conclude that a VHO procedure within MIH and/or ANDSF could take one of the following Forms:

*VHO Procedure1 includes ANDSF, FAF and a VHO algorithm.*
*VHO Procedure2 includes ANDSF, FAF, DFF and a VHO algorithm.*
*VHO Procedure3 includes ANDSF, MIH and a VHO algorithm.*
*VHO Procedure4 includes MIH and VHO algorithm.*

**Procedure (1)** requires FAF as one additional entity for two reasons: first, to enable the transmission from WiMAX to 3GPP and second, to avoid direct link between 3GPP and WiMAX. **Procedure (2)** requires two additional entities (FAF and DFF) in order to provide seamless VHO integrated with ANDSF. **Procedure (3)** includes the combination between two mechanisms (MIH and ANDSF) in order to provide seamless VHO without additional entities. To the best of our knowledge, no evaluations or validations have been presented, it is not exhaustive and it is complex as a result of combining between two mechanisms. In **Procedure (4),** utilizing MIH and present smart VHO algorithm will achieve four main goals: **a)** avoiding the complexity due to combination between MIH and ANDSF mechanisms, and avoiding any new additional entities (FAF and/or DFF) **b)** providing VHO with better performance **c)** less connection failure and **d)** more exhaustive for enhancing VHO heterogeneous wireless networks. This will be shown in I AM 4 VHO approach section.

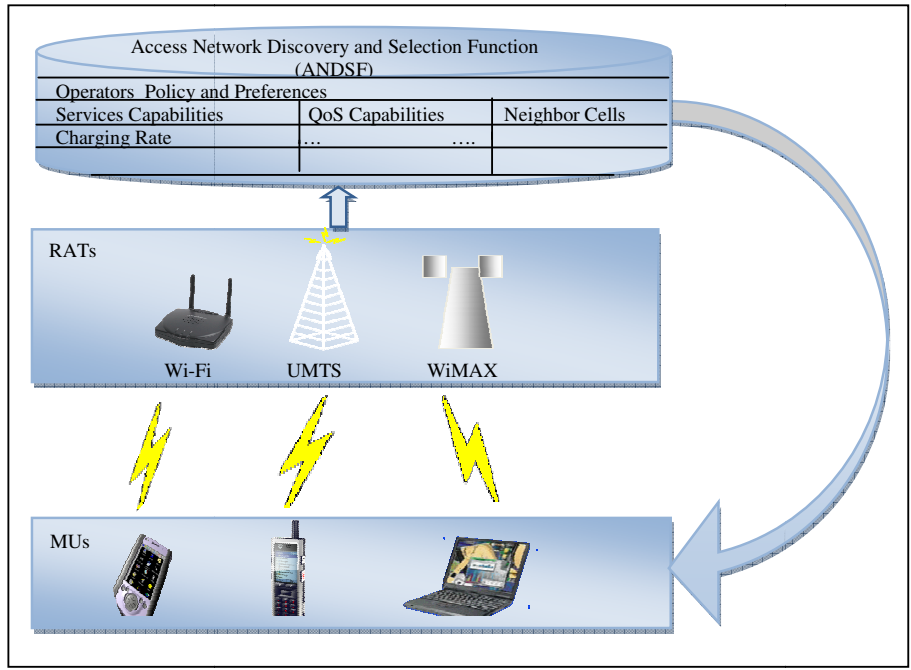

Figure 3: Access Network Discovery and Selection Function (ANDSF) passing information about Radio Access Technologies (RATs) to Mobile Users (MUs)

## 4. I AM 4 VHO APPROACH

In the literature there are three approaches which have been evolved from each other to provide seamless VHO [3, 11 and 12]. In [11], the authors lacked to tackle two vital aspects in the VHO procedure: first, the source network was not informed by the MU about its movement to the target network which resulted in packet losses and second, it lacked releasing procedure for the resources of the network. In [12], new additional entities were required (FAF and DFF) in order to provide seamless VHO. In [3], the authors proposed the combination between MIH and ANDSF mechanisms in order to provide seamless VHO without additional entities; however, it is not exhaustive and it is complex as a result of combining between two mechanisms. In this paper, we present new approach that can provide better performance, less connection failure, less complexity and more exhaustive compared to that in literature. Our approach consists of a procedure which is implemented by an algorithm and will provide the following:

1. Details on network operation in case of VHO initiated imperatively due to Radio Signal Strength (RSS) or alternatively due to the user preferences, such as low cost and high data rate, taking into account higher priority to execute imperative session (more exhaustive).
2. VHO algorithm based on our approach achieves less connection failure (probability of reject sessions) as a result of using the RATs list of priority. When the first choice from the RATs list of priority could not be satisfied with available resources, the Admission Control (AC) at destination PoS will automatically move to another RAT selection in the list in order to satisfy the requirements of this RAT selection and so on.
3. Better VHO performance with more soft (minimal packet loss) and more faster (minimal latency) due to start buffering new data packets that comes from Correspondent Node (CN) server after the RAT has been checked by destination PoS.
4. No need to combine between ANDSF and MIH mechanisms as in [3] as a result of assigning the operator's policies and preferences from PoS at the destination network (less complex).
5. Dynamic Host Configuration Protocol (DHCP) server to distribute the Care of Address (CoA) to mitigate the load on PoS.

### 4.1 I AM 4 VHO Procedure

We describe our proposed procedure through the VHO phases: initiation, decision and execution, this is shown in Fig.4. Three VHO scenarios are considered: UMTS to Wi-Fi, Wi-Fi to WiMAX and WiMAX to UMTS.

**Initiation Phase**

In this phase, while the MU is connected to the source network the VHO will mainly trigger imperatively due to RSS going down or alternatively based on the user preferences, e.g. high data rate and low cost.

**Decision Phase**

In this phase, as a result of mainly triggering in the initiation phase, *MIIS Request/Response Available RATs* message will be responsible to pass available RATs to MU via source network (PoA and PoS) such that MIIS is collocated with Home Agent (HA) [2,13], in imperative session due to RSS going down the MU will select RATs list of priority based on best reading RSS and then pass them to the destination PoS via source network, whereas in alternative session the MU will select RATs list of priority based on user preferences. When the first choice from the RATs list of priority could not be satisfied with available resources, the AC at destination PoS will automatically move to another RAT selection in the list in order to satisfy the requirements of this RAT selection and so on, once RAT resources has been found, it will be checked by destination PoS whether it is compliant with the rules and preferences of operators. If that is available, the MIIS/HA will be informed to start buffering new data packets which are sent by CN server.

**Execution Phase**

In this phase, The MU will be connected to target RAT to start its authentication with destination PoA and obtain CoA from DHCP. After that, *Update/Acknowledge binding* message notifies HA about new the CoA to start sending the buffered data and continuing the session within target network. Finally, after completion of sending the buffered data, the resources of the source network are released by MIH.

After the RAT has been checked by destination PoS, concurrent notification informs both of MIIS/HA server to start buffering and source PoS to pass selected target RAT to source PoA; after that, the source PoA sends the target RAT to MU for handover. The MU makes use of data buffering period in MIIS/HA server to start/end authentication messages with the destination PoA to obtain CoA, whereas the old data packets are still sent to the MU from CN server at the old IP address. After that, Update/Acknowledge binding message notifies HA about the new CoA to start sending the buffered data and continuing the session within the target RAT. This will achieve following: **a)** reduced time interval in which the MU does not receive any packets as a result of handover (latency) and **b)** low packet loss ratio.

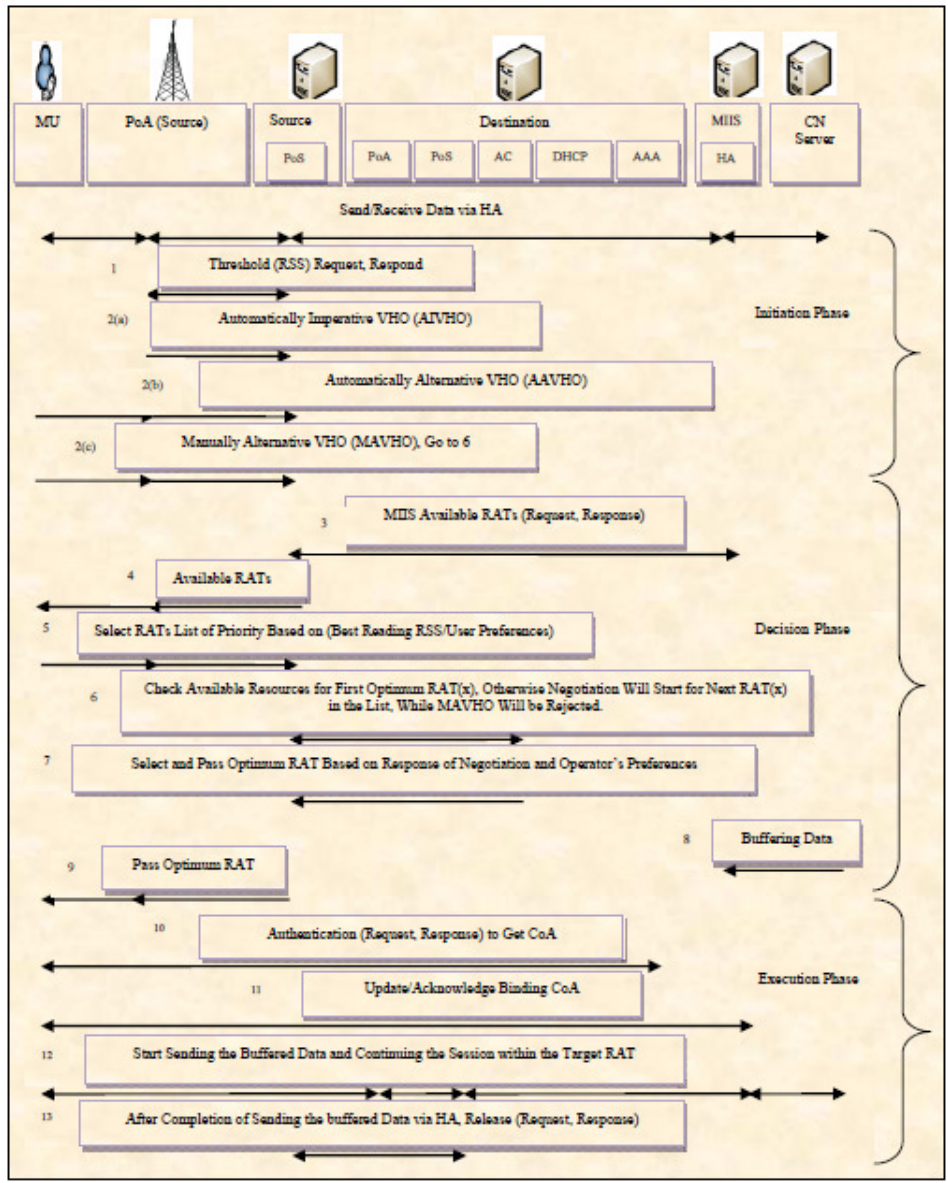

Figure 4: Imperative Alternative MIH for Vertical Handover procedure (I AM 4 VHO procedure)

## 4.2 I AM 4 VHO Algorithm

The algorithm to implement our proposed procedure defines two main types of VHO: Automatically Imperative VHO (AIVHO) session and Alternative VHO session (AVHO). The AVHO consists of Automatically Alternative VHO (AAVHO) session and Manually Alternative VHO (MAVHO) session, this is shown in Fig.5.

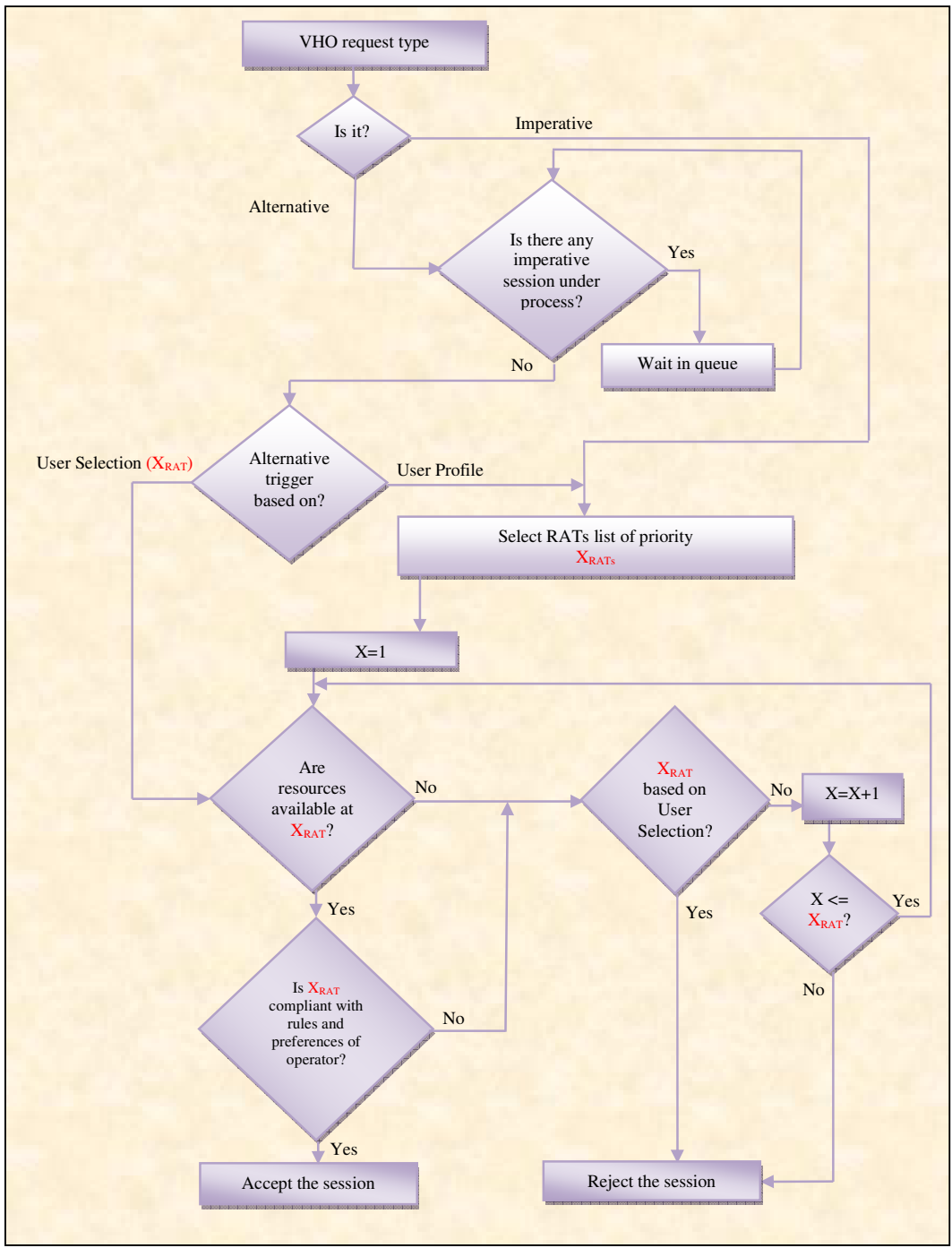

Figure 5: Imperative Alternative MIH for Vertical Handover algorithm (I AM 4 VHO algorithm)

Imperative session will have high priority, e.g. if there are two VHO sessions at the same time, one due to RSS going down (imperative) and the other due to user preferences change (alternative), the first request will be responded as high priority and the second request will be

61

considered only if there is no any imperative VHO session under process, otherwise it has to wait in queue. In the AIVHO case, due to RSS going down the RATs list of priority based on best reading RSS will be provided by MU. When the first choice from the RATs list of priority could not be satisfied with Sufficient of Resources (SoRs), the AC at destination PoS will automatically move to the next RAT in the list for satisfying the request and so on, once RAT of sufficient resources has been found, it will be checked by the destination PoS whether it is compliant to the rules and preferences of operators, if that is available, the session will be accepted, otherwise the request will be returned to the AC step to select the next RAT in list. Finally, the session will be rejected if there are no available resources for any RAT in the list. In the AAVHO case, the MU will select optimum RATs list of priority based on user profile change (e.g. data rate, low cost) and take the same path of imperative request. In the MAVHO case, there is no need to RATs list of priority step because the RAT is selected manually by the user; therefore, the session would be rejected if SoRs are not available for user's selection session.

## 5. CONCLUSION

In this paper, we have presented our I AM 4 VHO approach based on the approaches that have been studied in the literature with better performance (packet loss and latency), less connection failure (probability of reject sessions), less complexity and more exhaustive for enhancing VHO compared with similar approaches in the literature. It consists of a procedure which is implemented by an algorithm. The proposed VHO procedure is primarily based on MIH mechanism to compliment its work. Also, two types of VHO are defined and giving priority to imperative sessions over alternative sessions. In future work, we plan to simulate the proposed VHO approach and obtain results about system performance by considering different scenarios.

## REFERENCES

[1] Our vision: Committed to connecting the world. (04/08/2011). International Telecommunication Union (ITU).Retrieved 5 Jan, 2012, from http://www.itu.int/en/about/Pages/vision.aspx.

[2] Angoma, B.; Erradi, M.; Benkaouz, Y.; Berqia, A.; Akalay, M.C.; , "HaVe-2W3G: A vertical handoff solution between WLAN, WiMAX and 3G networks," *7th International Wireless Communications and Mobile Computing Conference (IWCMC)*, 4-8 Jul 2011, pp.101-106.

[3] Frei, S.; Fuhrmann, W.; Rinkel, A.; Ghita, B.V.; , "Improvements to Inter System Handover in the EPC Environment," *4th IFIP International Conference on New Technologies, Mobility and Security (NTMS)*,7-10 Feb 2011, pp.1-5.

[4] Zekri, M.; Jouaber, B.; Zeghlache, D.;,"Context aware vertical handover decision making in heterogeneous wireless networks," *IEEE 35th Conference on Local Computer Networks (LCN)*,10-14Oct 2010, pp.764-768.

[5] Kassar, M.; Kervella, B.; Pujolle, G.;,"An overview of vertical handover decision strategies in heterogeneous wireless networks," *Computer Communications*, vol.31, no.10, 25 Jun 2008, pp. 2607-2620.

[6] Louta, M.; Zournatzis, P.; Kraounakis, S.; Sarigiannidis, P.; Demetropoulos, I.;,"Towards realization of the ABC vision: A comparative survey of Access Network Selection," *IEEE Symposium on Computers and Communications(ISCC)*, 28 Jun 2011-1 Jul 2011, pp.472-477.

[7] IEEE 802.21Tutorial. (2006). IEEE802.21. Retrieved 5 Jan, 2013, from http://www.ieee802.org/21/.

[8] Lampropoulos, G.; Salkintzis,A.K.;Passas,N.;,"MediaIndependent Handover for Seamless Service Provision in Heterogeneous Networks," *IEEE Communication Magazine*, vol. 46, no. 1, Jan. 2008, pp.64-71.

[9] Xiaohuan,Y.;Y.Ahmet Şekercioğlu.;Sathya,N.;,"A survey of vertical handover decision algorithms in Fourth Generation heterogeneous wireless networks," *Computer Networks*,vol.54,no.11, 2 Aug 2010,pp.1848-1863.

International Journal of Computer Networks & Communications (IJCNC) Vol.5, No.3, May 2013

[10] Yahiya,T.A.; Chaouchi,H.;,"On the Integration of LTE and Mobile WiMAX Networks," *Proceedings of 19th International Conference on Computer Communications and Networks (ICCCN)* ,2-5Aug 2010, pp.1-5.

[11] Taaghol, P.; Salkintzis, A.; Iyer, J.; , "Seamless integration of mobile WiMAX in 3GPP networks," *IEEE Communications Magazine*,vol.46,no.10 ,Oct 2008, pp.74-85.

[12] Song, W.; Jong-Moon,C.; Daeyoung,L.; Chaegwon, L.; Sungho, C.; Taesun,Y.; , "Improvements to seamless vertical handover between mobile WiMAX and 3GPP UTRAN through the evolved packet core," *IEEE Communications Magazine* ,vol.47,no.4, Apr 2009, pp.66-73.

[13] Neves, P.; Soares, J.; Sargento, S.;,"Media Independent Handovers: LAN, MAN and WAN Scenarios," *IEEE GLOBECOM Workshops*, 30 Nov 2009 –4 Dec 2009, pp.1-6.

[14] Marquez-Barja, J.; Calafate, C.T.; Cano, J.-C.; Manzoni, P.;,"Evaluation of a technology-aware vertical handover algorithm based on the IEEE 802.21 standard," IEEE Wireless Communications and Networking Conference (WCNC), 28-31Mar 2011, pp.617-622.

[15] Capela, N.; Soares, J.; Neves, P.; Sargento, S.; , "An Architecture for Optimized Inter-Technology Handovers: Experimental Study," IEEE International Conference on Communications (ICC), 5-9 Jun 2011, pp.1-6.

[16] Heecheol,S.; Jongjin,K..; Jaeki,L.; Hwang,S.L.;,"Analysis of vertical handover latency for IEEE 802.21-enabled Proxy Mobile IPv6," 13th International Conference on Advanced Communication Technology (ICACT), 13-16 Feb 2011, pp.1059-1063.

**AUTHORS**

**Mr. Omar Khattab** is a Ph.D. student at the Department of Computing, Science and Engineering at Salford University, UK. He received MSc in Computer Information System, BSc in Computer Science and Diploma in Programming Language 2005, 2003 and 2000 respectively. He has seven years of teaching experience in Networks, Computer Science and Information Technology. He has obtained a lot of international professional certificates in the field computer networks. He is Microsoft Certified Trainer.
His current research area is Computer Networking and Data Telecommunication.

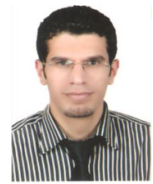

**Dr. Omar Alani** received his PhD degree in Telecommunication Engineering from De Montfort University, UK in 2005. He is currently a lecturer of telecommunications at the School of Computing, Science & Engineering, University of Salford in the UK. His research interests include Radio resource management and location/mobility management in next generation mobile communication systems, diversity and adaptive modulation techniques as well as Ad hoc networks.

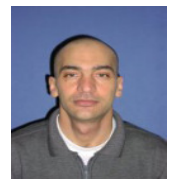